# Superconductivity to 262 kelvin via catalyzed hydrogenation of yttrium at high pressures

Elliot Snider[1,5], Nathan Dasenbrock-Gammon[2,5], Raymond McBride[1], Xiaoyu Wang[3], Noah Meyers[1], Keith V. Lawler[4], Eva Zurek[3], Ashkan Salamat[5], Ranga P. Dias[1,2]*

[1]Department of Mechanical Engineering, School of Engineering and Applied Sciences, University of Rochester, Rochester, NY 14627, USA.

[2]Department of Physics and Astronomy, University of Rochester, Rochester, NY 14627, USA.

[3]Department of Chemistry, State University of New York at Buffalo, Buffalo, NY, USA.

[4]Department of Chemistry & Biochemistry, University of Nevada Las Vegas, Las Vegas, Nevada 89154, USA

[5]Department of Physics & Astronomy, University of Nevada Las Vegas, Las Vegas, Nevada 89154, USA.

*These authors contributed equally.

Room temperature superconductivity has been achieved under high pressure in an organically derived carbonaceous sulfur hydride with a critical superconducting transition temperature ($T_c$) of 288 kelvin[1]. This development is part of a new class of dense, hydrogen-rich materials with remarkably high critical temperatures. Metal superhydrides are a sub-class of these materials that provide a different and potentially more promising route to very high-$T_c$ superconductivity. The most promising binary metal superhydrides contain alkaline or rare earth elements, and recent experimental observations of $LaH_{10\pm\delta}$ have shown them capable of $T_c$'s up to 250 to 260 kelvin[2,3]. Predictions have shown yttrium superhydrides to be the most promising with an estimated $T_c$ in excess of 300 kelvin for $YH_{10}$[3–6]. Here we report **the synthesis of an yttrium superhydride that exhibits superconductivity at a critical temperature of 262 kelvin at 182 ± 8 gigapascal. A palladium thin film assists the synthesis by protecting the sputtered yttrium from oxidation and promoting subsequent hydrogenation[7]. Phonon-mediated superconductivity is established by the observation of zero resistance, an isotope effect and the reduction of $T_c$ under an external magnetic field. The upper critical magnetic field is ~103 tesla at zero temperature. We suggest $YH_{9\pm\delta}$ is the synthesized product based on comparison of the measured Raman spectra and $T_c$ to calculated Raman results.**

Metallic hydrogen is considered to have a high Debye temperature and strong electron-phonon coupling that is necessary for high $T_c$ phonon-mediated superconductivity[8–11]. Based on the properties of metallic hydrogen[12,13], metal hydride compounds have long been investigated as a means toward increasing $T_c$ via the hydrogen acting as a "superconducting glue" providing high frequency vibrations[14], but early metal hydride studies were unable to achieve $T_c$'s over 20 K[15]. The discovery of superconductivity in hydrogen sulfide at 203 K changed the notion of what might be possible for phonon-mediated superconductors[16–19]. The resultant paradigm shift has cemented the understanding that a hydride compound must have a large electron-phonon coupling related to the hydrogens and a high density of hydrogen related states at the Fermi level (only achieved by forms of hydrogen other than $H_2$) to have a high $T_c$[5]. High hydrogen content materials that meet those two criteria can be considered as chemically pre-compressed phases relative to pure hydrogen, thus lowering the pressures necessary for metallization or high $T_c$ superconductivity to more experimentally feasible conditions[20].

High $T_c$ hydrogen-rich materials can be broadly categorized into two types: purely covalent metallic hydrides, such as hydrogen sulfide and the newly discovered room-$T_c$ carbonaceous sulfur hydride[1], and the so-called metal 'superhydrides'. In the metal superhydrides, metal atoms occupy the center of the cages of clathrate-like hydrogen lattices[4–6,21–25]. Predictions for many metal superhydrides are exceptionally encouraging for achieving unprecedented high $T_c$'s – the theoretical work on the lanthanum and yttrium superhydrides thus far has suggested $T_c$ values of 274–286 K at 210 GPa and reaching 305–326 K at 250 GPa, respectively[4–6,21–25]. Interestingly, scandium (the lightest rare earth) has been predicted to exhibit a range of superhydride stoichiometries with superconducting properties and exotic hydrogen bonding networks that are unique to those of Y and La, indicating an ion-size effect. Somewhat counterintuitively, many metal superhydrides with a lighter ion and more hydrogen present (e.g. $LiH_x$)[26] exhibit a lower $T_c$ than La and Y superhydrides. This is opposed to $YH_{10}$ and $MgH_6$ having a higher $T_c$ than $LaH_{10}$ and $CaH_6$, respectively. Although in both cases the lighter ion required higher pressures to reach the maximum $T_c$ values. The cause of these trends is related to the bonding of hydrogen in the superhydride, but the exact nature of the periodic trends remains to be seen.

The theoretical descriptions of the rare earth superhydrides have flourished since the initial discovery of superconducting sulfur hydride, providing exceptionally high estimates for $T_c$ in both La and Y superhydrides. While $LaH_{10}$ has been realized, the synthesis of yttrium superhydrides has lagged behind and experimental confirmations of the predicted near-ambient-temperature superconductivity have remained elusive. Herein, we present an alternative technique to overcome many of the hurdles inhibiting the synthesis of metal superhydrides in a diamond anvil cell. Using the technique of coating the target metal (yttrium) in a thin film of palladium – a metal known to promote hydrogenation, we report the discovery of superconductivity in an yttrium superhydride up to a maximum $T_c$ of 262 K at 182 ± 8 GPa. The observed Raman spectra and $T_c$ suggest a material with stoichiometry close to $YH_9$ is responsible for the observed superconductivity based on comparison to previous structure search and Bardeen–Cooper–Schrieffer theory simulations[4,5,21–23].

Yttrium and the trivalent rare-earth elements are extremely reactive, which makes them difficult to study[27]. To overcome these difficulties, we employed a non-conventional synthesis technique in a diamond anvil cell (DAC) — sputtering an yttrium film (400 to 800 nm) in the middle of the diamond culet under high vacuum and coating it with a thin layer of palladium (10 ± 5 nm). Palladium is a noble metal with a well-known ability to catalyze hydrogenation, hydrogen purification, and hydrogen storage by promoting uptake of hydrogen into its lattice then transferring it to other materials[28]. Palladium accomplishes this by dissociating molecular hydrogen and incorporating the hydrogen atoms into octahedral sites of the fcc Pd lattice ($\beta$-PdH$_x$). The pre-dissociation of molecular hydrogen by palladium thin films along with its high hydrogen transport[29], has already been shown to catalyze yttrium and lanthanum metals into REH$_{3-\delta}$ at 300 K and $0.9\times10^5$ Pa of H$_2$ in a matter of minutes[7]. Using our setup (Fig. 1a) akin to that of Huiberts et al.[7], we were able to fully transform a sputtered yttrium sample to YH$_3$ in a DAC over a period of ~18 hours at 4.5 GPa at room temperature without any other external stimuli[30]. This procedure ensures that our starting material for a high pressure yttrium superhydride synthesis is YH$_3$. Starting with the trihydride is advantageous over other approaches as it provides a higher initial hydrogen content than direct elemental combination, and it avoids using chemical precursors as a source of hydrogen (e.g. ammonia borane) which very likely leads to contamination beyond a pure binary system.

To synthesize the yttrium superhydride, the dense, film-coated YH$_x$ (x≥3) and H$_2$ are compressed above 130 GPa and directly heated by a V-GEN Yb Pulsed Fiber Laser at 1064 nm (Fig. 1b). The estimated lower bound of the temperature of the samples during heating is ~1800

K. Details of the laser-heating experiments have been described elsewhere[35]. The Pd should be stoichiometric (or near) β-PdH by 100 GPa, and laser heating removes all barriers to hydrogen diffusion through β-PdH$_x$[31]. Thus, we believe that the laser heated Pd is again acting to catalyze the formation of the superhydride by removing the barriers associated with dissociating molecular H$_2$ and driving the transport of H to the yttrium system. Thus, this method permits a facile transformation pathway for metal superhydride materials predicted to have high $T_c$'s.

The synthesized yttrium superhydride has a maximum superconducting transition temperature of 262 K at 182 ± 8 GPa, as is evident by a sharp drop in resistance over a 5 to 10 degree temperature change (Fig. 1c). Furthermore, all the yttrium superhydride samples prepared using this technique exhibit a high $T_c$ immediately post laser heating at high pressure. The transition temperature increases with pressure from 216 K at 134 ± 5 GPa until it plateaus around 175 GPa, which is then followed by a slight decrease approaching a dome shape (see Fig. 1c inset). At 144 GPa, the superconducting transition is signified by two sharp resistance drops: a very large (~75%) drop that occurs relatively slowly ($\Delta T = 8$ K) at around 244 K, and a small (~25%) but sharp ($\Delta T < 1$ K) drop at around 237 K (indicated by * in the Fig. 1C). A similar behavior is also observed at the highest pressure measured, 182 ± 8 GPa, with the 262 K transition. These pressures are measured using the H$_2$ vibron scale[32], but it should be noted that pressures measured from the diamond edge using the Akahama 2006 scale[33] are at least 10% to 12% higher. Also, the transitions temperatures were taken from the onset of superconductivity using a probe with ± 0.1 K accuracy, where the resistance is measured during the natural warming cycle (~0.25 K/min) from low temperature with a current of 10 μA to 1 mA.

A pronounced shift of $T_c$ from an isotopic substitution would suggest that the synthesized yttrium superhydrides superconduct via an electron–phonon mechanism consistent with the Bardeen–Cooper–Schrieffer (BCS) theory of conventional superconductors. To investigate the isotope effect, $D_2$ replaced $H_2$ in the experimental setup and yttrium superdeuteride samples were synthesized at the same conditions as the superhydrides. The Raman spectrum of the yttrium superdeuteride obtained in such a fashion agrees reasonably well with $YD_9$ at 175 GPa (Fig. S12), complementing the spectroscopic assignment of the superhydride discussed below. The substitution of deuterium noticeably affects the value of the $T_c$ which shifts to lower temperatures, indicating phonon-assisted superconductivity (Fig. 2). The isotopic shift in critical temperature is expected to scale as $\propto m^{-\alpha}$, with $m$ being the atomic mass and $\alpha$ the isotope coefficient. A strong isotope effect is observed at 177 GPa with $T_c = 256$ K for yttrium superhydride and $T_c = 183$ K for the superdeuteride. The isotope coefficient obtained from those measurements is $\alpha = -\,[\ln T_c\,(YD_x) - \ln T_c\,(YH_x)]/\ln 2 = 0.48$, in very good agreement with the Bardeen–Cooper–Schrieffer value of $\alpha \approx 0.5$ for conventional superconductivity.

Our typical yttrium superhydride samples are about 30–50 μm in diameter, making it almost impossible to search for the Meissner effect in the DC magnetization or even the shielding effect in the DC or AC magnetic susceptibility[2,3]. These difficulties signal the need for novel experimental capabilities[34–37], but an alternative method to confirm a superconducting transition at high pressure exploits the inherent hostility of an external magnetic field on superconductivity. The upper critical field, $H_c$, is the maximum external magnetic field that superconductivity can survive. This value is controlled by both the paramagnetic effect of electron spin polarization due to the Zeeman effect and the diamagnetic effect of orbital motion of the Cooper pairs due to the Lorentz force. Under an applied magnetic field, these effects combine to reduce the $T_c$. In the present studies, application of an external magnetic field reduces $T_c$ by about 14 K at 6 T and 177 GPa (Fig. 3), confirming the superconducting transition. The upper critical field, $H_c(T) =$ $H_c(0)[1 - \left(\frac{T}{T_c}\right)^2]$, in the 0 K limit is 103.2 T (Fig. 3, top inset). The field dependence of the superconducting transition width, $\Delta T_c$, at 177 GPa is linear (Fig. 3, bottom inset) as is expected from the percolation model[33]. $\Delta T_c$ is defined here as $\Delta T_c = T_{90\%} - T_{10\%}$ where $T_{90\%}$ and $T_{10\%}$ are the temperatures corresponding to 90% and 10% of the resistance at 260 K at 177 GPa. Fitting to $\Delta T_c = \Delta T_c(0) + kH$ gives a transition width at zero external field $\Delta T_c(0)$ = 7.62 K and proportionality constant $k$ = 0.07 KT$^{-1}$. Such a large $\Delta T_c(0)$ indicates sample inhomogeneities, which is typical for DAC experiments. For instance, the transition width for LaH$_{10}$ is ~ 7.3 K and H$_3$S can vary from 5.5 K to 26 K[3,16,38].

There have been moderate differences between the high $T_c$ responses of the RE superhydrides reported[2,3,39,40]. Most relevant is our higher $T_c$ of nearly 20 K, compared to Kong et al[39]. The potential of mixed phasing present in either this study or any of the other metal superhydride phases reported cannot be excluded. However, the superconducting response measured is most likely from a single phase that is above the necessary phase fraction for bulk conductivity measurements. It is important to note that the abundancy of the phase fractions will be defined by the percolation of the new thermodynamically stable phase and any kinetic barriers on the pathway. There are several possible reasons other than mixed phasing for discrepancies of superconducting temperatures from different synthesis procedures including the use of a chemical precursor for a source of hydrogen vs. elemental starting materials or the specific reaction pathway. Alternatively, the synthesized superhydride materials could have defects in the lattice where H differs from the ideal stoichiometry, ie. YH$_{9\pm\delta}$. This is a known issue in lower H-content yttrium hydrides, for instance the formation of YH$_{2.86}$ at near ambient pressures in the earlier Pd film work[7]. The alignment of grains at boundaries are also known to affect the performance of ceramic superconductors[41]. Such subtle changes in these systems must be probed by spectral techniques beyond diffraction that are sensitive to the local atomistic coordination environments.

Raman scattering is a spectral technique that is highly sensitive to the bonding environment of materials, and figure 4 shows Raman spectra that were obtained before and after pulsed heating of a mixture of dense $YH_x$ (x≥3) and $H_2$ at 171 GPa. The changes in the spectra signify that the sample transforms to an entirely new structure upon laser heating. Before heating the Raman spectra of the sample is featureless, and the spectrum post-heating has three sharp, characteristic Raman peaks around 505, 833, and 1136 cm$^{-1}$ and a broad peak at ~2048 cm$^{-1}$. The broad peak contains a shoulder at ~1890 cm$^{-1}$ (Fig. 4a). After several cycles of pulsed laser heating, a reasonable pressure drop (1–2% of the initial pressure) is observed. The drop suggests a volume collapse is associated with the transition to the superhydride.

There is a wealth of theoretical results available to identify the possible structures for yttrium superhydrides at very high pressures[4–6,19,21–24]. The most probable predicted stoichiometries with high superconducting transition temperatures are $YH_6$, $YH_9$, and $YH_{10}$. Comparison of the measured Raman spectrum at 187 GPa (Figure S11a top) with the ones computed for various $YH_n$ phases at 190 GPa (Fig. S9b) immediately shows that the experimental results cannot be explained by Im-3m $YH_6$ (because its only Raman active mode is within the diamond zone), nor $P6_3/m$ symmetry $YH_9$ (because it has an intense peak at 3630 cm$^{-1}$ due to the $H_2$ vibron). Considering that the computational methodology employed is likely to underestimate the vibrational frequencies by ~240 cm$^{-1}$, the $YH_{10}$ mode computed to lie at 1703 cm$^{-1}$ could potentially account for the broad shoulder observed at ~2200 cm$^{-1}$. However, $YH_{10}$ only has two modes whose frequencies fall below that of the diamond mode, whereas experiment observes three centered at 582, 899, and 1221 cm$^{-1}$ (Figure S11a, top).

Turning to P6$_3$/mmc YH$_9$, we observe good agreement between the calculated and observed spectrum at three different pressures (Figure S11a), especially considering that within this pressure range considered our computational methodology underestimated the H$_2$ vibron by ~243 cm$^{-1}$. We postulate that the broadening due to quantum nuclear effects is particularly important for the high frequency modes above 1500 cm$^{-1}$, which could merge into a single peak with a pronounced shoulder similar to what is observed experimentally. A comparison of the experimental and theoretical frequencies of the major peaks vs. pressure (Figure S11c) plots is also in good agreement, especially taking into account that the theoretical results are likely too low. We also calculated the Raman spectrum of P6$_3$/mmc YD$_9$ at 180 GPa and compared the results with experiment (Figure S12 bottom). The calculated and measured peak positions are in excellent agreement with each other, however we overestimate the intensity of the mode centered at 949 cm$^{-1}$. The pronounced differences in the computed spectra of these phases illustrate that Raman can serve as a spectral fingerprint unveiling the motifs present in the hydrogenic sublattice.

Note that these calculations are only indicative, as the mode frequencies are sensitive to the choice of the exchange correlation function and anharmonic effects[42]. Upon further compression, that low frequency mode vanishes potentially signaling the creation of YH$_{10}$. However, the exceptionally high $T_c$ predicted for YH$_{10}$ was not measured. Our results show that we have successfully synthesized an yttrium superhydride material at high pressure and elevated temperatures which we tentatively assign as YH$_{9\pm\delta}$.

In conclusion, Pd catalyzed hydrogenation of sputtered yttrium produced a yttrium superhydride with a maximum superconducting transition temperature of 262 K at 182 ± 8 GPa. Analysis using Raman spectroscopy reveals a uniform material with similar features to YH$_9$ but with some subtle differences, leading us to determine that we have synthesized YH$_{9\pm\delta}$. We believe this method of delivering hydrogen to the sample will permit stoichiometric tuning of these superhydride materials thus a tuning of the superconducting temperature. Despite extensive efforts, we do not observe room-temperature superconductivity up to the highest pressure studied, therefore the search for RTSC in these metal superhydrides remains open. Based on the strong isotope effect, the mechanism of pairing is convincingly conventional, but further research is needed for a complete understanding of the underlying mechanism in this new class of quantum materials — metal superhydrides. The synthesis of very high $T_c$ superconductors that are stable (or metastable) at ambient pressure remains a grand challenge but with both new experimental techniques and robust theoretical tools now available, a path is open to tailoring superhydrides to "materials by design" approaches in extreme science for transformative technologies.

**Figure Captions:**

**Fig. 1 | Superconductivity in yttrium superhydride at high pressures. a,** Schematic of new experimental approach. Two methods are used to prepare the yttrium film. In the first method, a film of yttrium is deposited in the middle of the culet of one diamond, with a thickness of 400 nm to 800 nm and a diameter of 70 μm to 80 μm[43]. The yttrium is protected from oxidation by a deposited palladium layer of 10 ± 5 nm thick. In the second method, a yttrium foil with an initial thickness of ~5 μm is squeezed to approximately 1 μm thick film between diamond anvils; the 1 μm film with diameter of ~60 μm is placed on one of the diamonds in the experimental cell with protected Pd layer. To further prevent oxidation of the metal surface, the sample and diamond anvil cell were handled in a glove box charged with inert gas. Hydrogen gas (>99.999% purity) is loaded into the cell to form $YH_3$ either by means of a high-pressure gas loading device at ~0.3 GPa at room temperature or cryogenically at 14 K. The loaded hydrogen serves both as the source for uptake by the yttrium film and as a pressure medium. Both diamonds are coated with ~40 nm of alumina to prevent hydrogen from diffusing into the diamonds. To synthesize yttrium superhydride, mixtures of dense $YH_x$ (x≥3) + $H_2$ are directly heated at above 130 GPa with a pulsed laser of wavelength 1064 nm. The estimated lower bound of the temperature of the samples is ~1800 K across 21 experiments. Based on our extensive experience, the alumina coating does not affect or contaminate the sample at high pressures and even at temperatures as high as ~2500 K[12,13]. The possibility of $PdH_x$ being the source of the observed superconductivity has been investigated and discussed extensively in the supplementary materials[43]. **b,** Microphotographs from transmitted light of yttrium film in a hydrogen environment. (i) The yttrium film is initially opaque to the visible spectrum at ambient conditions. (ii) When yttrium is exposed to $H_2$ gas, it readily absorbs hydrogen and forms metallic stoichiometric dihydride ($YH_{2±ε}$) and, eventually, nonmetallic, hcp

trihydride ($YH_{3-\delta}$ with $\delta \simeq 0.2$) at low pressure[43]. The mixture of $YH_{3-\delta}$ and $H_2$ is further pressurized (about 5 GPa), fully transparent $YH_3$ is synthesized. The trihydride is transparent and yellowish, with an optical gap of 2.6 eV[43]; however, the transmission shifts toward the red (iii and iv) above 15 GPa, and eventually the sample again becomes completely opaque to the visible spectrum (v) above 60 GPa at ambient temperature, providing visual evidence for the metallization of $YH_3$. The bottom right image (vi) illustrates the synthesized superhydride sample ($YH_x$: $x \geq 6$ + $H_2$) with electrical leads in a four-probe configuration for resistance measurements. **c,** Temperature-dependent electrical resistance of yttrium superhydride at high pressures, showing the superconducting transitions as high as 262 K at 182 ± 8 GPa, the highest pressure measured in this experimental run. The data were obtained during the warming cycle to minimize the electronic and cooling noise. **Inset:** The pressure dependence of the $T_c$ as determined by sharp drop in electrical resistance, showing the increase in $T_c$ with pressure and plateau around 175 GPa, approaching a dome shape. The colors represent different experiments.

**Fig. 2 | The Isotope effect.** The substitution of deuterium noticeably affects the value of the $T_c$, which shifts to 183 K at 177 GPa. The calculated isotope coefficient at 177 GPa with $T_c$ = 256 K for **$YH_{9\pm\delta}$** and $T_c$ = 183 K for **$YD_{9\pm\delta}$** sample is 0.48. The green and orange curve represents **$YH_{9\pm\delta}$** and **$YD_{9\pm\delta}$** respectively. By comparing the transition temperatures at around 183 GPa, we obtained $\alpha$ = 0.46. Both values are in very good agreement with the Bardeen–Cooper–Schrieffer value of $\alpha \approx 0.5$ for conventional superconductivity.

**Fig. 3 | Superconducting transition under an external magnetic field.** Low-temperature electrical resistance behavior under magnetic field of at H = 0 T, 1 T, 3 T, and 6 T (increasing from right to left) at 177 GPa. Pressure was determined by hydrogen vibron position. **Inset top:** Upper critical field versus temperature. An extrapolation to the lowest temperature yields ~103.2 T for the upper critical magnetic field in the limit of zero temperature. **Inset bottom:** Superconducting transition width, $\Delta T_c$, as a function of $H$, showing a linear relationship.

**Fig. 4 | Synthesis of yttrium super-hydride at high pressures and high temperatures. a,** Pressure-induced Raman changes (background subtracted) of yttrium superhydride for several indicated pressures at room temperature. Before heating (bottom spectra), the pressure was at 171 GPa; after heating, it dropped to 168 GPa, suggesting a volume collapse associated with the transition. The observed characteristic Raman modes of the new material are in good agreement with theoretical calculations for $YH_9$[43]. The peak around 2048 cm$^{-1}$ is broad compared to the low-frequency peaks, and there may be a shoulder at ~1890 cm$^{1}$. The shaded area corresponds to the strong diamond signal. **Inset:** Microphotographs of transmitted light of the sample before and after heating. For clarity we have included a line (yellow) around the boundary of the sample. The sample became darker after heating to above 1800 K. **b,** Pressure-induced Raman shift of different vibrations of the new yttrium superhydride. The colors indicate different experiments.

**Acknowledgments**

We thank Hiranya Pasan for reanalyzing the low temperature data, and Desmond Wentling for his technical support during the initial stage of the project. Also, we thank Ori Noked, Christian Koelbl, and Lauren Koelbl for useful discussions. Preparation of diamond surfaces was performed in part at the University of Rochester Integrated Nanosystems Center. Calculations were performed at the Center for Computational Research at SUNY Buffalo [44]. This research was supported by NSF, Grant No. DMR-1809649, and DOE Stockpile Stewardship Academic Alliance Program, Grant No. DE-NA0003898. This work supported by the U.S. Department of Energy, Office of Science, Fusion Energy Sciences under Award Number DE-SC0020340. AS and KVL are supported by DE-FOA-0002019. XW and EZ are supported by DE-SC0020340 and NSF, Grant No. DMR-1827815.

**Competing interests**

The authors declare no competing interests

**Data and materials availability**

The data reported in this paper are tabulated and available upon reasonable request from R.P.D.

**Corresponding author**

Correspondence should be addressed to Ranga P. Dias, rdias@rochester.edu, (585) 276 4112

**Fig. 1 | Superconductivity in yttrium superhydride at high pressures.**

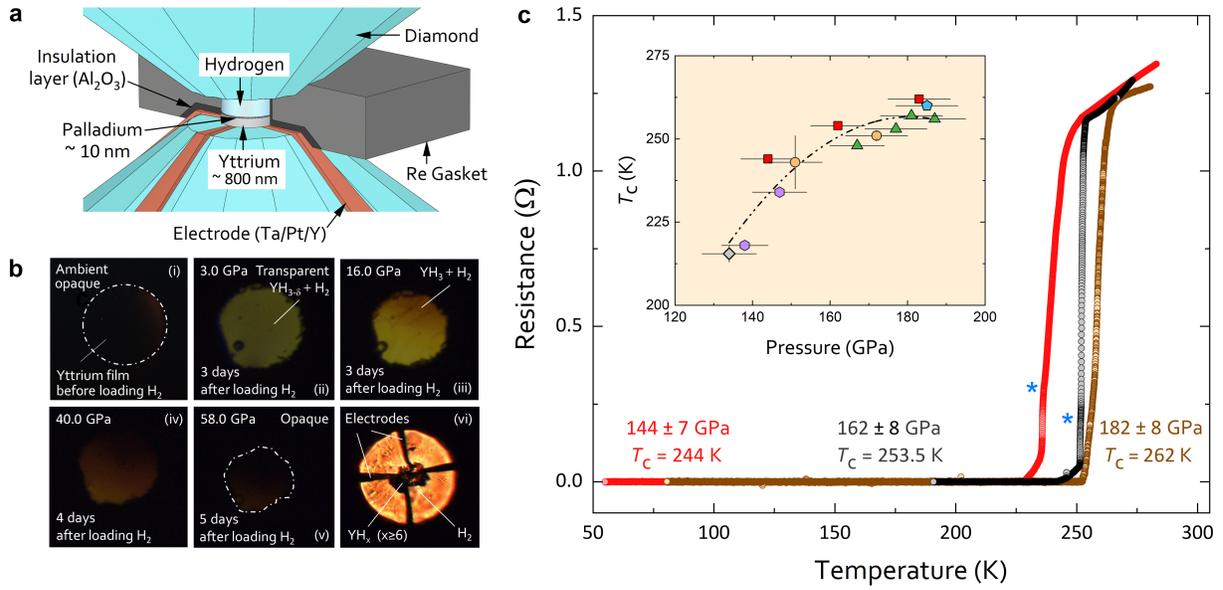

**Fig. 2 | The Isotope effect.**

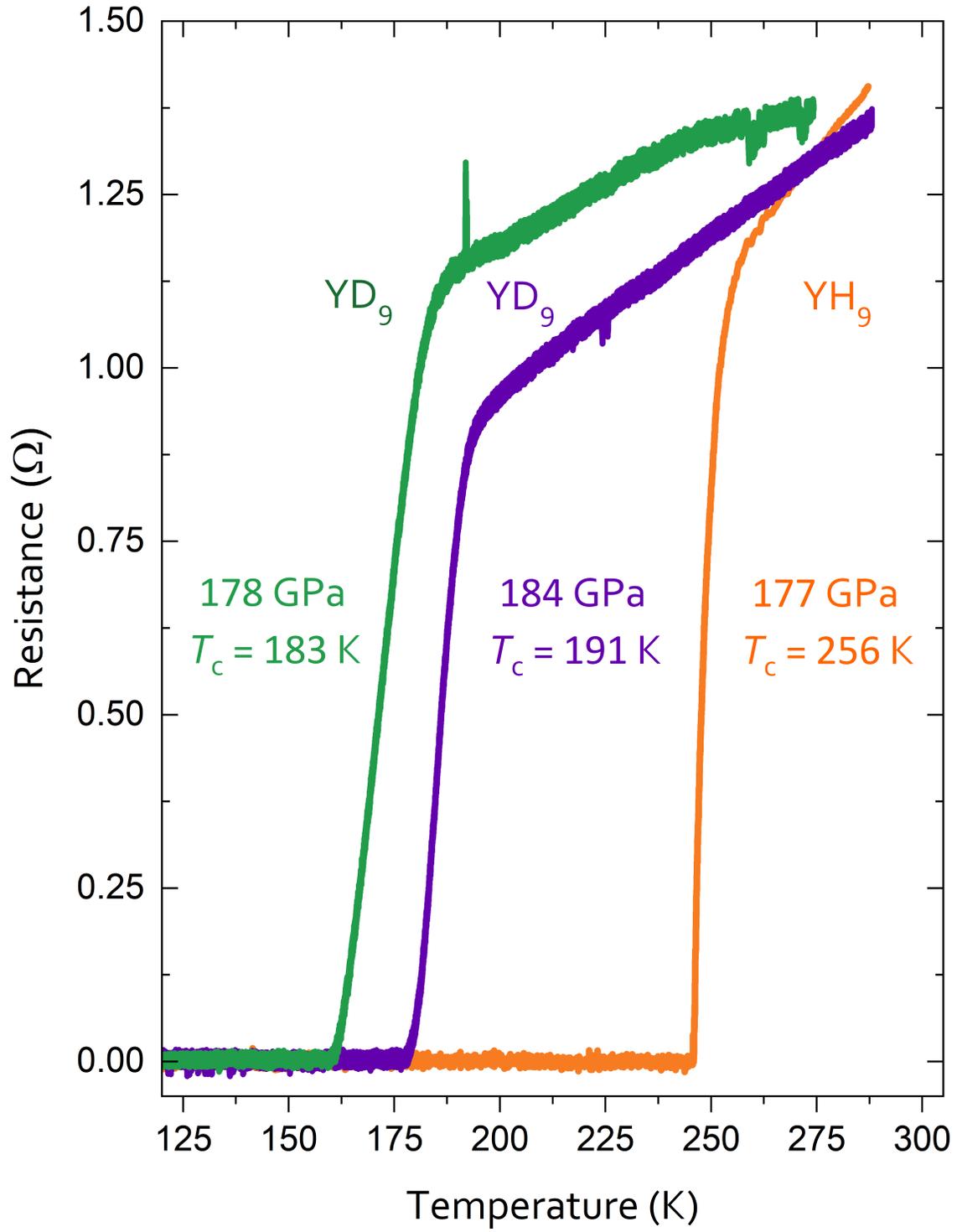

**Fig. 3 | Superconducting transition under an external magnetic field.**

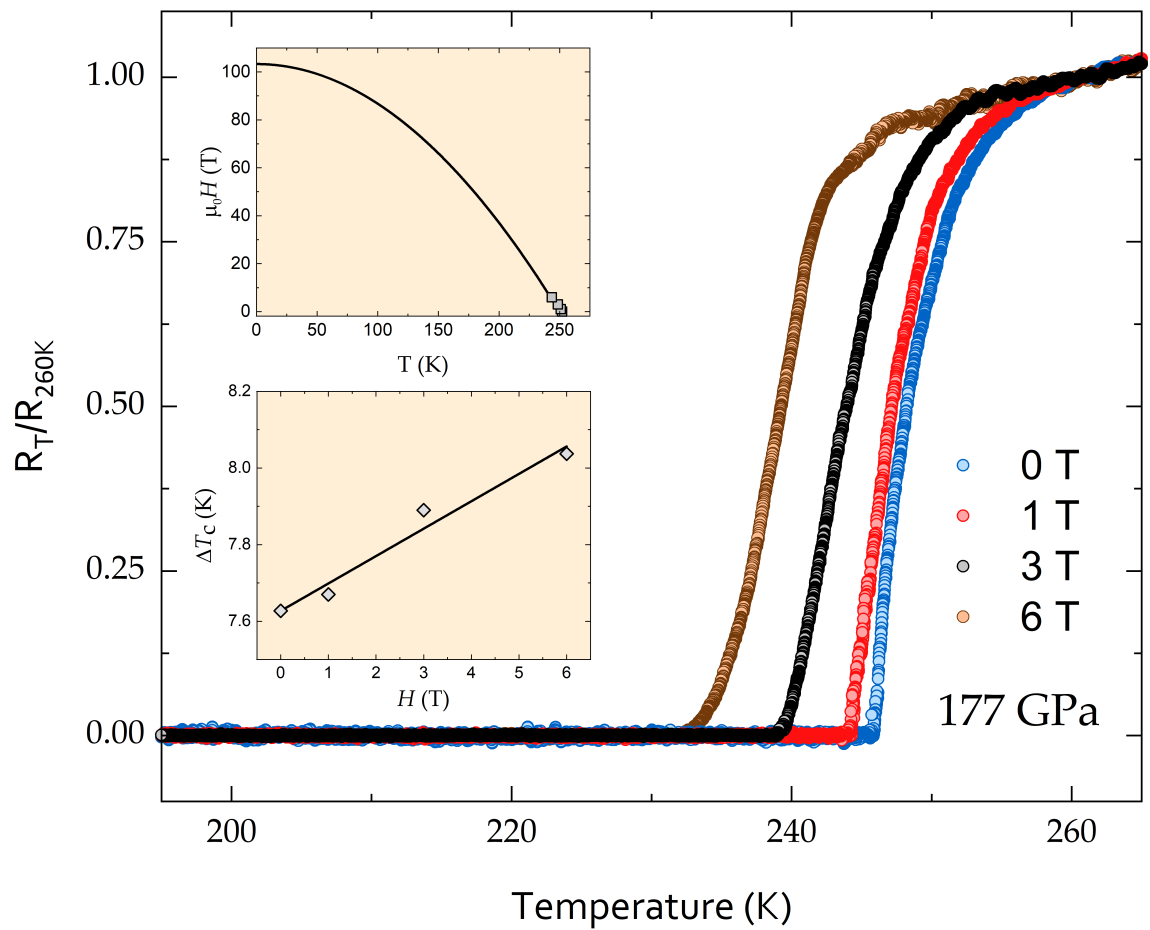

**Fig. 4 | Synthesis of yttrium super-hydride at high pressures and high temperatures.**

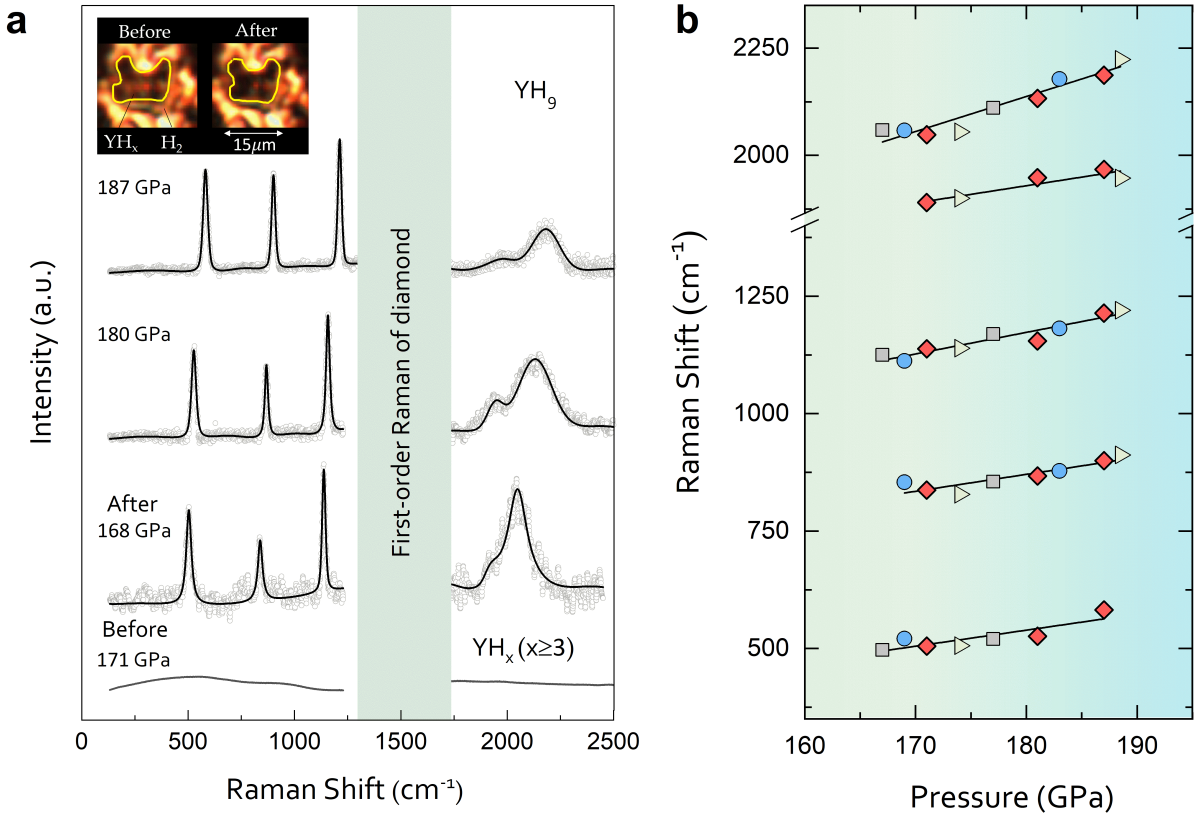